\documentclass[aps,preprint,amsmath,amssymb,superscriptaddress,nofootinbib]{revtex4}
\usepackage{multirow}
\usepackage{graphicx} 
\usepackage{amsmath}
\usepackage{float}
\begin{document}

\title{Sensitivity analysis for the anomalous $W^+W^-\gamma$ couplings at the LHeC and the FCC-he}

\author{S. Spor}
\email[]{serdar.spor@beun.edu.tr}
\affiliation{Department of Medical Imaging Techniques, Zonguldak B\"{u}lent Ecevit University, 67100, Zonguldak, Turkey.}

\author{M. K{\"o}ksal}
\email[]{mkoksal@cumhuriyet.edu.tr} 
\affiliation{Department of Physics, Sivas Cumhuriyet University, 58140, Sivas, Turkey.} 

\begin{abstract}
We study the anomalous $W^+W^-\gamma$ couplings in performing the production of the W-pair through the process $e^- p\,\rightarrow\,e^-\gamma^*\gamma^* p\,\rightarrow\,e^-W^+W^-p$ at the Large Hadron electron Collider (LHeC) and the Future Circular Collider-hadron electron (FCC-he). We examine the pure hadronic, the semi-leptonic and the pure leptonic decay channels of the produced W-pair. We assume the center-of-mass energies of $\sqrt{s}=1.30$ and $1.98$ TeV and the integrated luminosities of 10-100 fb$^{-1}$ for the LHeC and the center-of-mass energies of $\sqrt{s}=3.46$ and $5.29$ TeV and the integrated luminosities of 100-2000 fb$^{-1}$ for the FCC-he. The best limits at 95$\%$ confidence level obtained for the anomalous couplings $\Delta\kappa_{\gamma}$ and $\lambda_{\gamma}$ are $\Delta\kappa_{\gamma}=[-0.00208; 0.00206]$ and $\lambda_{\gamma}=[-0.00174; 0.00834]$. By comparing these limits with the experimental limits, we present the potentials of the process and future $ep$ colliders.
\end{abstract}

%\pacs{12.60.-i}

\maketitle

\section{Introduction}

The Standard Model (SM) is a very powerful particle physics theory for predicting the properties and interactions of elementary particles. It has also been tested in many important experiments and has proven to be quite successful, especially after the discovery of a particle consistent with the Higgs boson. However, some fundamental questions remain unanswered and new physics beyond the SM is expected to find answers to these problems. New physics effects can be investigated by examining the anomalous gauge boson interactions determined by non-Abelian $\text{SU}_\text{L}(2)\times\text{U}_\text{Y}(1)$ gauge symmetry. The most accurate investigation of these gauge couplings can provide valuable information for new physics beyond the SM, testing the gauge structure of the SM and the spontaneous symmetry breaking mechanism. Any deviations of the trilinear gauge boson couplings from expected values will be key elements in the search for new physics beyond the SM.

Potential deviations from the SM predictions due to new physics for the anomalous triple gauge couplings (aTGC) in the electroweak sector can be parameterized in a model-independent framework through the effective Lagrangian method. This method reveals the low energy effects of new physics to be found at higher energies. The Lagrangian includes new operators built from the SM fields and invariant under its symmetries. These operators are larger in dimension than four and are suppressed by the negative forces of the new physics scale $\Lambda$. The effective Lagrangian considering the higher dimension operators is given as

\begin{eqnarray}
\label{eq.1} 
{\cal L}_{eft}={\cal L}_{SM}+\sum_i\frac{c_i^{(6)}}{\Lambda^2}{\cal O}_i^{(6)}+\sum_j\frac{c_j^{(8)}}{\Lambda^4}{\cal O}_j^{(8)}+\cdots
\end{eqnarray}

{\raggedright where ${\cal L}_{SM}$ is SM Lagrangian, ${c_i^{(6)}}$ and ${c_j^{(8)}}$ are the couplings of dimension-six operators of ${\cal O}_i^{(6)}$ and dimension-eight operators of ${\cal O}_j^{(8)}$ and $\Lambda$ is new physics scale.} 

In the evaluation of $WW\gamma$ couplings, $SU\left(2\right) \times U\left(1\right)$ invariant dimensional-6 operators that contribute to the possible new physics beyond the SM are used and the effective lagrangian of the new physics generated by these operators is given as follows \cite{Baak:2013yre}:

\begin{eqnarray}
\label{eq.2} 
{\cal L}_{eft}=\frac{1}{\Lambda^2}\left[C_W{\cal O}_W+C_B{\cal O}_B+C_{WWW}{\cal O}_{WWW}+C_{\tilde{W}WW}{\cal O}_{\tilde{W}WW}+C_{\tilde{W}}{\cal O}_{\tilde{W}}+h.c.\right]\,,
\end{eqnarray}

{\raggedright where the CP-conserving dimension-six operators ${\cal O}_{WWW}$, ${\cal O}_{W}$ and ${\cal O}_{B}$ and the CP-violating dimension-six operators ${\cal O}_{\tilde{W}WW}$ and ${\cal O}_{\tilde{W}}$ are}

\begin{eqnarray}
\label{eq.3} 
{\cal O}_{WWW}=\text{Tr}\left[W_{\mu\nu}W^{\nu\rho}W_\rho^\mu\right]\,,
\end{eqnarray}
\begin{eqnarray}
\label{eq.4} 
{\cal O}_{W}=\left(D_\mu\Phi\right)^\dagger W^{\mu\nu}\left(D_\nu\Phi\right)\,,
\end{eqnarray}
\begin{eqnarray}
\label{eq.5} 
{\cal O}_{B}=\left(D_\mu\Phi\right)^\dagger B^{\mu\nu}\left(D_\nu\Phi\right)\,,
\end{eqnarray}
\begin{eqnarray}
\label{eq.6} 
{\cal O}_{\tilde{W}WW}=\text{Tr}\left[\tilde{W}_{\mu\nu}W^{\nu\rho}W_\rho^\mu\right]\,,
\end{eqnarray}
\begin{eqnarray}
\label{eq.7} 
{\cal O}_{\tilde{W}}=\left(D_\mu\Phi\right)^\dagger \tilde{W}^{\mu\nu}\left(D_\nu\Phi\right)\,.
\end{eqnarray}

Here $\Phi$ is the Higgs doublet field, $D_\mu$ is the covariant derivative and $B^{\mu\nu}$ and $W^{\mu\nu}$ are field strength tensors of $SU\left(2\right)_L$ and $U\left(1\right)_Y$ gauge groups, respectively. The anomalous $WW\gamma$ coupling is parametrized in terms of the Lagrangian \cite{Hagiwara:1987xdj,Bian:2015ylk}:

\begin{eqnarray}
\label{eq.8} 
\begin{split}
{\cal L}_{WW\gamma}=& \,ig_{WW\gamma}\Big[{g_1^{\gamma}}\left({W_{\mu\nu}^{+}}{W_{\mu}^{-}}A_{\nu}-{W_{\mu\nu}^{-}}{W_{\mu}^{+}}A_{\nu}\right) \\
&+\kappa_{\gamma}{W_{\mu}^{+}}{W_{\nu}^{-}}A_{\mu\nu}+\frac{\lambda_{\gamma}}{M_W^2}{W_{\mu\nu}^{+}}{W_{\nu\rho}^{-}}A_{\rho\mu}  \\
&+ig_4^\gamma {W_{\mu}^{+}}{W_{\nu}^{-}}\left( \partial_\mu A_\nu+\partial_\nu A_\mu\right) \\
&-ig_5^\gamma \epsilon_{\mu\nu\rho\sigma}\left({W_{\mu}^{+}}\partial_\rho{W_{\nu}^{-}}-\partial_\rho{W_{\mu}^{+}}{W_{\nu}^{-}}\right)A_\sigma \\
&+\tilde{\kappa}_{\gamma}{W_{\mu}^{+}}{W_{\nu}^{-}}\tilde{A}_{\mu\nu}+\frac{\tilde{\lambda}_{\gamma}}{M_W^2}{W_{\lambda\mu}^{+}}{W_{\mu\nu}^{-}}\tilde{A}_{\nu\lambda} \Big]\,,
\end{split}
\end{eqnarray}

{\raggedright where $g_{WW\gamma}=-e$ and $\tilde{A}=\frac{1}{2}\epsilon_{\mu\nu\rho\sigma}A_{\rho\sigma}$. The field strength tensors $A^{\mu\nu}=\partial^\mu A^\nu - \partial^\nu A^\mu$ and $W_{\mu\nu}^\pm=\partial_\mu W_\nu^\pm - \partial_\nu W_\mu^\pm$ are defined for photon and $W$ bosons. Electromagnetic gauge invariance requires that ${g_1^{\gamma}}=1$ and $g_4^\gamma=g_5^\gamma=0$. The first three terms of above Eq.~(\ref{eq.8}) conform to $C$ and $P$ while the other four terms violate $C$ and/or $P$. However, there are two $C$ and $P$-conserving parameters: $\kappa_{\gamma}$ and $\lambda_{\gamma}$; and two $C$ and/or $P$-violating parameters: $\tilde{\kappa}_{\gamma}$ and $\tilde{\lambda}_{\gamma}$. In the SM at tree level, the anomalous coupling values are $\kappa_{\gamma}=1$ ($\Delta \kappa_{\gamma}=0$) and $\lambda_{\gamma}=0$.}

In this study, $\kappa_{\gamma}$ and $\lambda_{\gamma}$ couplings that are $C$ and $P$-conserving have been focused on. The couplings of the Lagrangian in Eq.~(\ref{eq.8}) can be written in terms of the couplings of the Lagrangian in Eq.~(\ref{eq.2}) via $\kappa_{\gamma}=1+\Delta \kappa_{\gamma}$ as \cite{Hagiwara:1993aso,Wudka:1994abs,Degrande:2013rry}

\begin{eqnarray}
\label{eq.9} 
{\Delta \kappa_\gamma}=\left(c_W+c_B\right)\frac{m_W^2}{2\Lambda^2}\,,
\end{eqnarray}
\begin{eqnarray}
\label{eq.10} 
{\lambda_\gamma}=c_{WWW}\frac{3g^2m_W^2}{2\Lambda^2}\,.
\end{eqnarray}

Phenomenological studies for the anomalous ${WW\gamma}$ coupling have been performed on the LHC \cite{Sahin:2011dfg,Etesami:2016eto,Bian:2015ylk,Bian:2016wer,Falkowski:2015ghe,Chapon:2010thw,Bhatia:2019gso,Sahin:2017uot}, the LHeC \cite{Mariotto:2012yur,Cakir:2014mjp,Li:2018tyb,Rodriguez:2020erv,Koksal:2020opr}, the ILC \cite{Kumar:2015ghe,Spor:2020rpo,Rahaman:2020abs}, the CEPC \cite{Bian:2015ylk,Bian:2016wer}, the CLIC \cite{Ari:2016aac,Billur:2019suu} and the FCC-he \cite{Rodriguez:2020erv,Koksal:2020opr}. The experimental limits at 95$\%$ C.L. on $\Delta \kappa_{\gamma}$ and $\lambda_{\gamma}$ couplings were reported by the CMS \cite{Sirunyan:2019umc}, ATLAS \cite{Aaboud:2017les}, CDF \cite{Aaltonen:2010tyf}, D0 \cite{Abazov:2012mwc}, ALEP, DELPHI, L3, OPAL \cite{Schael:2013msh}, see Table~\ref{tab1}. Furthermore, the phenomenological limits were obtained in studies using future collider parameters and some of these limits are given in Table~\ref{tab1}.

\begin{table}[H]
\caption{The experimental and phenomenological limits at 95$\%$ C.L. on $\Delta \kappa_{\gamma}$ and $\lambda_{\gamma}$ couplings from the present and future colliders.}
\label{tab1}
\centering
\begin{ruledtabular}
\begin{tabular}{ccc}
Experimental limit & $\Delta \kappa_{\gamma}$ & $\lambda_{\gamma}$\\ 
\hline
CMS Collaboration \cite{Sirunyan:2019umc} & [-0.0275; 0.0286] & [-0.0065; 0.0066]\\
ATLAS Collaboration \cite{Aaboud:2017les} & [-0.0610; 0.0640] & [-0.0130; 0.0130]\\
CDF Collaboration \cite{Aaltonen:2010tyf} & [-0.5700; 0.6500] & [-0.1400; 0.1500]\\
D0 Collaboration \cite{Abazov:2012mwc} & [-0.1580; 0.2550] & [-0.0360; 0.0440]\\
ALEP, DELPHI, L3, OPAL \cite{Schael:2013msh} & [-0.0990; 0.0660] & [-0.0590; 0.0170]\\ 
\hline \hline
Phenomenological limit & $\Delta \kappa_{\gamma}$ & $\lambda_{\gamma}$\\ 
\hline
LHeC \cite{Cakir:2014mjp} & [-0.1820; 0.7930] & [-0.0390; 0.0790]\\
LHeC \cite{Li:2018tyb} & [-0.0030; 0.0021] & [-0.0034; 0.0021]\\
ILC \cite{Rahaman:2020abs} & [-0.0011; 0.0010] & [-0.0017; 0.0017]\\
ILC \cite{Baer:2013ttt} & [-0.00037; 0.00037] & [-0.00051; 0.00051]\\
CEPC \cite{Bian:2015ylk} & [-0.00045; 0.00045] & [-0.00033; 0.00033]\\
CEPC \cite{Ari:2016aac} & [-0.00102; 0.00103] & [-0.00168; 0.00173]\\
CLIC \cite{Billur:2019suu} & [-0.00007; 0.00007] & [-0.00004; 0.00102]\\
CLIC \cite{Ari:2016aac} & [-0.0004; 0.0023] & [-0.0007; 0.0007]\\
FCC-he \cite{Rodriguez:2020erv} & [-0.0013; 0.0013] & [-0.0046; 0.0046]\\
FCC-he \cite{Koksal:2020opr} & [-0.00069; 0.00069] & [-0.00990; 0.00540]\\
\end{tabular}
\end{ruledtabular}
\end{table}

\section{Cross-section of the process $e^- p\,\rightarrow\,e^-\gamma^*\gamma^* p\,\rightarrow\,e^-W^+W^-p$ at the LHeC and the FCC-he}

\begin{figure}[H]
\centering
\includegraphics[scale=0.61]{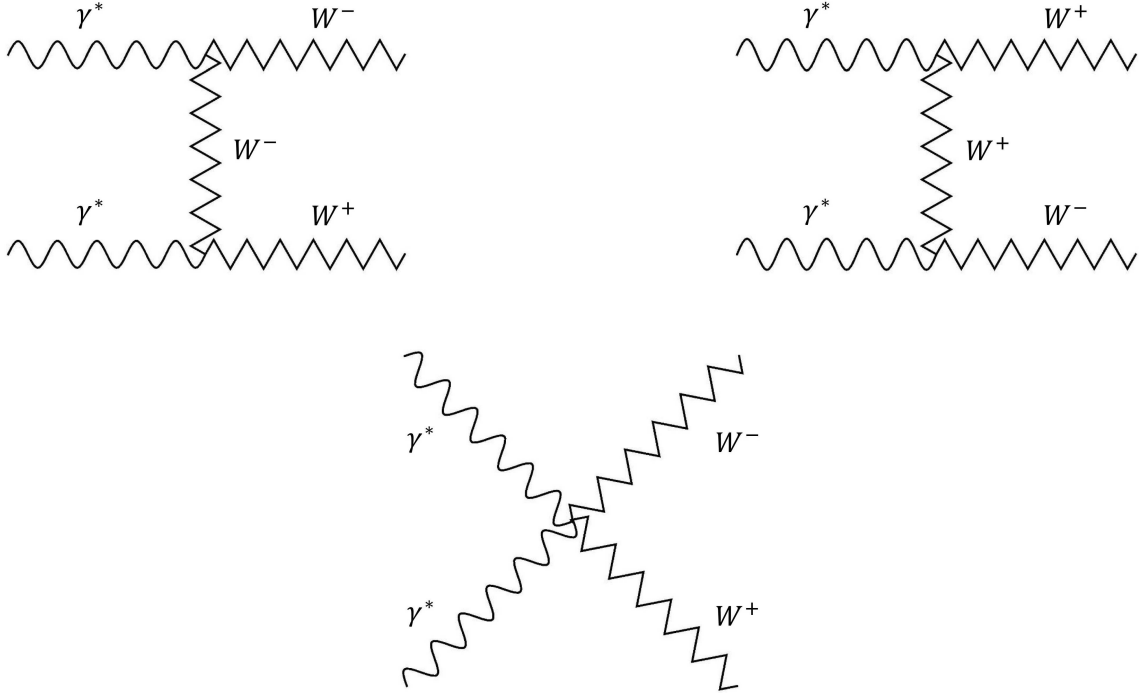}
\caption{The Feynman diagrams of the subprocess $\gamma^*\gamma^*\,\rightarrow\,W^+W^-$. 
\label{fig1}}
\end{figure}

We study the anomalous ${WW\gamma}$ couplings with $W^+W^-$ production in the process $e^- p\,\rightarrow\,e^-\gamma^*\gamma^* p\,\rightarrow\,e^-W^+W^-p$ at the LHeC and the FCC-he. The Feynmann diagrams for the subprocess $\gamma^*\gamma^* \,\rightarrow\,W^+W^-$ of the process $e^- p\,\rightarrow\,e^-\gamma^*\gamma^* p\,\rightarrow\,e^-W^+W^-p$ are given in Fig.~\ref{fig1}. The emitted $\gamma^*$ is a quasi-real photon scattered from electron or proton beams according to the framework of the Weizsacker-Williams Approximation (WWA), also known as the Equivalent Photon Approximation (EPA) \cite{Budnev:1975kyp,Baur:2002hjh,Piotrzkowski:2001tvz}. This quasi-real photons are scattered at very small angles from the beam pipe and have a low virtuality so that they can almost be assumed to be on mass-shell. The presence of these processes, which are called the photon-induced processes in the literature, has been proven by experiments at the Tevatron \cite{Abulencia:2007ghe,Aaltonen:2009yun,Aaltonen:2009erc,Abazov:2013pka} and the LHC \cite{Chatrchyan:2012oeq,Chatrchyan:2012mmn,Chatrchyan:2013yuw} and by many phenomenological studies \cite{Atag:2010hvb,Epele:2012ylm,Gupta:2012ezx,Sun:2014gds,Koksal:2016rei,Hernandez:2019aaa,Sahin:2020saw}.

In the EPA, two photons emitted from both electron and proton beams collide with each other in the subprocess $\gamma^*\gamma^* \,\rightarrow\,W^+W^-$. The spectrum of the first photon emitted from the electron is as follows \cite{Budnev:1975kyp}:

\begin{eqnarray}
\label{eq.11}
\begin{split}
f_{\gamma^{*}_{1}}(x_{1})=&\, \frac{\alpha}{\pi E_{e}}\Bigg\{\left[\frac{1-x_{1}+x_{1}^{2}/2}{x_{1}}\right]\text{log}\left(\frac{Q_{\text{max}}^{2}}{Q_{\text{min}}^{2}}\right)-\frac{m_{e}^{2}x_{1}}{Q_{\text{min}}^{2}}\left(1-\frac{Q_{\text{min}}^{2}}{Q_{\text{max}}^{2}}\right)
\\
&-\frac{1}{x_{1}}\left[1-\frac{x_{1}}{2}\right]^{2}\text{log}\left(\frac{x_{1}^{2}E_{e}^{2}+Q_{\text{max}}^{2}}{x_{1}^{2}E_{e}^{2}+Q_{\text{min}}^{2}}\right)\Bigg\}\,,
\end{split}
\end{eqnarray}

{\raggedright where $x_{1}=E_{\gamma_{1}^{*}}/E_{e}$. $Q_{\text{max}}^{2}$ is maximum virtuality of the photon. The minimum value of $Q_{\text{min}}^{2}$ is given as} 

\begin{eqnarray}
\label{eq.12}
Q_{\text{min}}^{2}=\frac{m_{e}^{2}x_{1}^{2}}{1-x_{1}}\,.
\end{eqnarray}

The spectrum of the second photon emitted from the proton is as follows \cite{Budnev:1975kyp,Rodriguez:2021hut}:

\begin{eqnarray}
\label{eq.13}
f_{\gamma^{*}_{2}}(x_{2})=\frac{\alpha}{\pi E_{p}}\left[1-x_{2}\right]\left[\varphi\left(\frac{Q_{\text{max}}^{2}}{Q_{0}^{2}}\right)-\varphi\left(\frac{Q_{\text{min}}^{2}}{Q_{0}^{2}}\right)\right]\,,
\end{eqnarray}

{\raggedright where the function $\varphi$ can be written by}

\begin{eqnarray}
\label{eq.14}
\begin{split}
\varphi(\theta)=&\left(1+ay\right)\left[-\textit{In}\left(1+\frac{1}{\theta}\right)+\sum_{k=1}^{3}\frac{1}{k(1+\theta)^{k}}\right]+\frac{y(1-b)}{4\theta(1+\theta)^{3}} \\
&+c\left(1+\frac{y}{4}\right)\left[\textit{In}\left(\frac{1-b+\theta}{1+\theta}\right)+\sum_{k=1}^{3}\frac{b^{k}}{k(1+\theta)^{k}}\right]\,.
\end{split}
\end{eqnarray}

{\raggedright In Eq.~(\ref{eq.14}), $y$, $a$, $b$ and $c$ parameters are expressed as follows:}

\begin{eqnarray}
\label{eq.15}
y=\frac{x_{2}^{2}}{1-x_{2}}\,,
\end{eqnarray}
\begin{eqnarray}
\label{eq.16}
a=\frac{1+\mu_{p}^{2}}{4}+\frac{4m_{p}^{2}}{Q_{0}^{2}}\approx 7.16\,,
\end{eqnarray}
\begin{eqnarray}
\label{eq.17}
b=1-\frac{4m_{p}^{2}}{Q_{0}^{2}}\approx -3.96\,,
\end{eqnarray}
\begin{eqnarray}
\label{eq.18}
c=\frac{\mu_{p}^{2}-1}{b^{4}}\approx 0.028\,.
\end{eqnarray}

The total cross section of the process $e^- p\,\rightarrow\,e^-\gamma^*\gamma^* p\,\rightarrow\,e^-W^+W^-p$ is determined by integrating over both the cross section of the subprocess $\gamma^*\gamma^* \,\rightarrow\,W^+W^-$ and the spectrums of two photons emitted from both electron and proton. The total cross section is as follows,

\begin{eqnarray}
\label{eq.19}
\sigma_{e^- p\,\rightarrow\,e^-\gamma^*\gamma^* p\,\rightarrow\,e^-W^+W^-p}=\int f_{\gamma^{*}_{1}}(x_{1})f_{\gamma^{*}_{2}}(x_{2}) d\hat{\sigma}_{\gamma^*\gamma^* \,\rightarrow\,W^+W^-} dx_{1} dx_{2}\,.
\end{eqnarray}

This numerical integration is performed using the CalcHEP package program in which the WWA vertex is embedded \cite{Belyaev:2013uhz}.

Investigating the cleanest and most sensitive parton dynamics in protons and nuclei, the deep inelastic lepton-hadron scattering (DIS) plays a decisive role in our understanding of the strong interaction, while also providing the exploratory potential to make precise measurements and reveal new particles \cite{Mellado:2013yqp,Caldwell:2018mvz}. Experimental data on the DIS were obtained by the HERA in DESY, the first $ep$ collider ever built. Between 1994 and 2007, it operated in colliding a 27.5 GeV electron beam with a 920 GeV proton beam. The center-of-mass energy was about $\sqrt{s}=320$ GeV and the total integrated luminosity was about 1 fb$^{-1}$ collected by the H1 and ZEUS Collaborations \cite{Abramowicz:2015tre}.

CERN designs two major future colliders for deeply inelastic lepton-hadron scattering; the Large Hadron electron Collider (LHeC) \cite{Fernandez:2012yub} and the Future Circular Collider-hadron electron (FCC-he) \cite{Abada:2019uqs,Bruning:2017pla}. The LHeC project aims to combine the proton beams of the Large Hadron Collider (LHC) with a new electron accelerator arranged tangentially to the main tunnel of the LHC. The LHeC is designed to collide electrons with an energy of 60 GeV to 140 GeV with protons of 7 TeV energy. Integrated luminosity ranging from 100 fb$^{-1}$ to 1 ab$^{-1}$ has been reported in design plans over the years. Similarly, as the LHeC and the LHC colliders perform together, the FCC-he and the Future Circular Collider-hadron hadron (FCC-hh) are designed to operate simultaneously. The FCC-he is planned with a 50 TeV proton beam from the FCC-hh and 60 GeV or higher electron beam. The FCC-he is expected to reach an integrated luminosity of 2-3 ab$^{-1}$.

The $ep$ collider has advantages over a $pp$ collider for some reasons. The $ep$ collider provides a clean environment with suppressed backgrounds from QCD strong interactions and it is also no effects of issues such as pileup, multiple interactions which reduce data quality in analysis. Besides, backward and forward scattering can be disentangled because the initial states are asymmetrical. Thus, it increases the signal significance and provides the opportunity to observe the phenomena that cannot be observed in a $pp$ collider. In this respect, an $ep$ collider will provide much needed complementary information to the LHC's physics program. It also provides additional and sometimes unique ways to study Higgs boson and physics beyond the SM, as well as top quark and electroweak physics \cite{Hesari:2018bxm}.

In this study, we consider two different stages in each of the LHeC and the FCC-he colliders. The center-of-mass energies of the LHeC with integrated luminosities in the range of 10 to 100 fb$^{-1}$ and the FCC-he with integrated luminosities in the range of 100 to 2000 fb$^{-1}$ are $\sqrt{s}=1.30$ and $1.98$ TeV and $\sqrt{s}=3.46$ and $5.29$ TeV, respectively, as two stages for each collider. The energy and integrated luminosity values used for the LHeC and the FCC-he are given in Table~\ref{tab2}.

\begin{table}[H]
\caption{The used values of electron-proton colliders.}
\label{tab2}
\centering
\begin{tabular}{p{3cm}p{2cm}p{2cm}p{2cm}p{4cm}}
\hline \hline
Colliders & $E_e$ (GeV) & $E_p$ (TeV) & $\sqrt{s}$ (TeV) & ${\cal L}_{\text{int}}$ (fb$^{-1}$)\\ 
\hline
LHeC-1 & 60 & 7 & 1.30 & 10, 30, 50, 100\\ 
LHeC-2 & 140 & 7 & 1.98 & 10, 30, 50, 100\\ 
FCC-he-1 & 60 & 50 & 3.46 & 100, 500, 1000, 2000\\ 
FCC-he-2 & 140 & 50 & 5.29 & 100, 500, 1000, 2000\\  \hline \hline
\end{tabular}
\end{table}

\begin{figure}[H]
\centering
\includegraphics[scale=1]{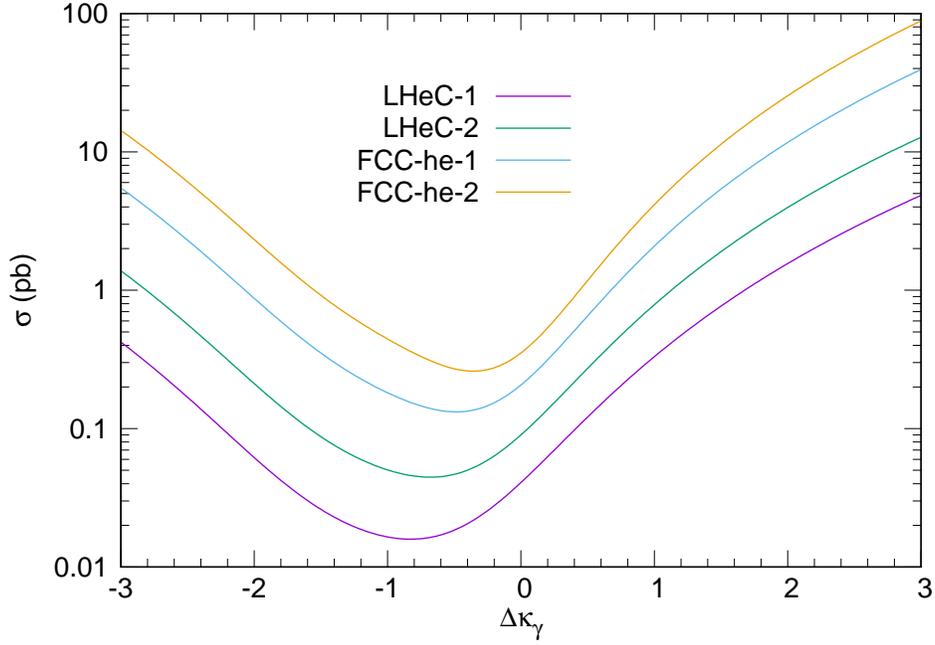}
\caption{The total cross sections of the process $e^- p\,\rightarrow\,e^-\gamma^*\gamma^* p\,\rightarrow\,e^-W^+W^-p$ as a function of the anomalous coupling $\Delta\kappa_{\gamma}$ at the LHeC and the FCC-he colliders. 
\label{fig2}}
\end{figure}

\begin{figure}[H]
\centering
\includegraphics[scale=1]{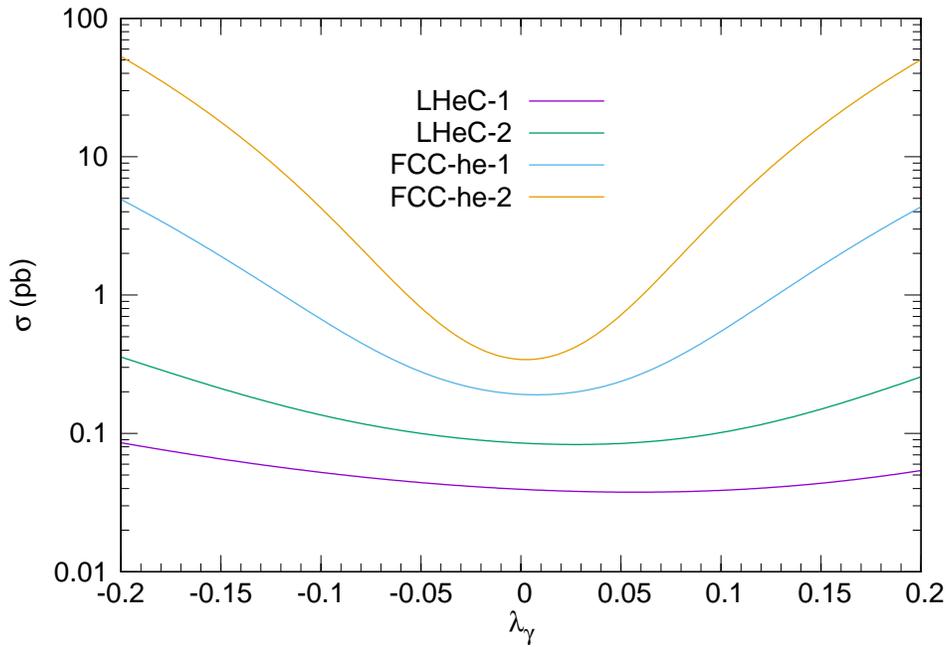}
\caption{Same as in Fig.~\ref{fig2}, but for the anomalous coupling $\lambda_{\gamma}$.     
\label{fig3}}
\end{figure}

The total cross sections of the process $e^- p\,\rightarrow\,e^-\gamma^*\gamma^* p\,\rightarrow\,e^-W^+W^-p$ according to the anomalous couplings $\Delta\kappa_{\gamma}$ and $\lambda_{\gamma}$ are shown in Figs.~\ref{fig2}-\ref{fig3}. The total cross sections corresponding to the range of $-3\leq \Delta\kappa_{\gamma} \leq3$ in Fig.~\ref{fig2} and $-0.2\leq \lambda_{\gamma} \leq 0.2$ in Fig.~\ref{fig3} are examined and the total cross sections at the FCC-he-2 collider reaches the highest values with the order of 90 pb and 45 pb, respectively. For both anomalous coupling parameters, it is noticed that as the center-of-mass energies of the LHeC and the FCC-he increase, the total cross sections also increase. There is an asymmetry of the total cross sections according to the negative and positive values of $\Delta\kappa_{\gamma}$ and $\lambda_{\gamma}$. As the center-of-mass energy of the $ep$ colliders increases, the minimum total cross sections shift to the right in the anomalous coupling $\Delta\kappa_{\gamma}$ and to the left in the anomalous coupling $\lambda_{\gamma}$.

\section{Bounds on the anomalous $\Delta\kappa_{\gamma}$ and $\lambda_{\gamma}$ couplings of the process $e^- p\,\rightarrow\,e^-\gamma^*\gamma^* p\,\rightarrow\,e^-W^+W^-p$ at the LHeC and the FCC-he}

We have studied 95$\%$ confidence level (C.L.) bounds on the anomalous couplings $\Delta\kappa_{\gamma}$ and $\lambda_{\gamma}$ using a $\chi^2$ analysis with systematic errors to determine the precision at the LHeC and the FCC-he. The $\chi^2$ function can be written as

\begin{eqnarray}
\label{eq.20} 
\chi^2=\left(\frac{\sigma_{SM}-\sigma_{NP}}{\sigma_{SM}\sqrt{\left(\delta_{st}\right)^2+\left(\delta_{sys}\right)^2}}\right)^2
\end{eqnarray}

{\raggedright where $\sigma_{SM}$ is the SM cross section and $\sigma_{NP}$ is the cross section containing contributions from both the SM and new physics beyond the SM. $\delta_{st}=\frac{1}{\sqrt {N_{SM}}}$ and $\delta_{sys}$ are the statistical error and the systematic error. The number of events in the SM is presented by $N_{SM}=S \times {\cal L}_{\text{int}} \times \sigma_{SM} \times BR$, where $S$ is the survival probability factor, ${\cal L}_{\text{int}}$ is the integrated luminosity and $BR$ is the branching ratio for W-pair decay.}

The hadronic nature of the proton should be comprised for photon-induced processes. This is done by adding a correction factor called the survival probability factor to the cross section. The survival probability factor is defined as the probability that scattered protons do not dissociate due to secondary soft interactions and it should be taken into account in a $\gamma^*$-proton or a $\gamma^*\gamma^*$ collision \cite{Sahin:2012opl,Senol:2015wcz}. In this study, we assume that the value of $S$ parameter is 0.7 for the $\gamma^*\gamma^*$ collision. We consider pure hadronic, semi-leptonic and pure leptonic decays of the W-pair production in the process $e^- p\,\rightarrow\,e^-\gamma^*\gamma^* p\,\rightarrow\,e^-W^+W^-p$. The branching ratios of W-pair decay are $BR\left(W^+W^-\rightarrow qqqq\right)=0.4544$ for pure hadronic, $BR\left(W^+W^-\rightarrow qq{\ell}\nu\right)=0.2877$ for semi-leptonic and $BR\left(W^+W^-\rightarrow {\ell}\nu{\ell}\nu\right)=0.0455$ for pure leptonic channels $\left({\ell}=e,\mu \right)$ \cite{Zyla:2020gmn}.

We present the sensitivity measurement for the anomalous couplings $\Delta\kappa_{\gamma}$ and $\lambda_{\gamma}$ at 95$\%$ C.L. without and with systematic errors 3$\%$, 5$\%$ in Tables~\ref{tab3}-\ref{tab6}. The bounds are obtained for pure leptonic, semi-leptonic and pure hadronic decay channels of the process $e^- p\,\rightarrow\,e^-\gamma^*\gamma^* p\,\rightarrow\,e^-W^+W^-p$ at the LHeC with $\sqrt{s}=1.30, 1.98$ TeV and ${\cal L}_{\text{int}}=10, 30, 50, 100$ fb$^{-1}$ and at the FCC-he with $\sqrt{s}=3.46, 5.29$ TeV and ${\cal L}_{\text{int}}=100, 500, 1000, 2000$ fb$^{-1}$.

The systematic uncertainties play an important role in the measurement of the anomalous couplings $\Delta\kappa_{\gamma}$ and $\lambda_{\gamma}$. The nature of the measurement apparatus, assumptions made by the experientialist, or the model used to make inferences based on the observed data are several reasons for the occurrence of systematic uncertainties. Therefore, the incorporation of systematic uncertainties in production of the W-pair is another remarkable aspect of our study. Some of the most common sources of systematic uncertainties are: luminosity, background and parton distribution functions and so on. In this work, we take into account the impact of total systematic uncertainties of 0, 3, 5$\%$ on the anomalous couplings $\Delta\kappa_{\gamma}$ and $\lambda_{\gamma}$.

As can be seen in Tables~\ref{tab3}-\ref{tab6}, although the luminosity on the couplings is increased due to the systematic error, the limits do not increase in proportion to the luminosity. With the improvement of the systematic error, it is inevitable that the limits on the couplings are better; e.g., the limits can be further improved by a factor of 4 at the projected luminosity with no systematic error. It can be said that the reason for this situation is the statistical error, which is much smaller than the systematic error.

\begin{figure}[H]
\centering
\includegraphics[scale=0.8]{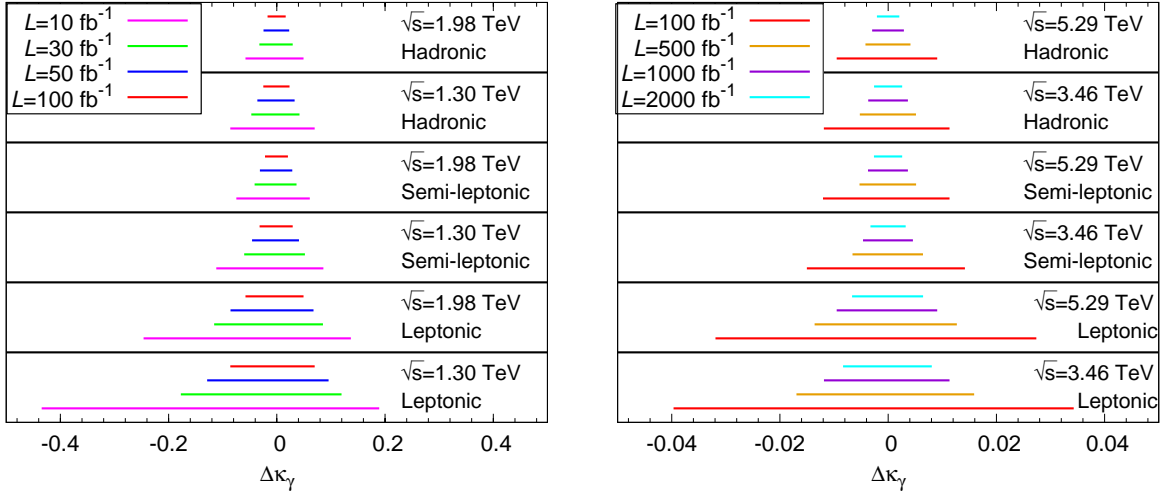}
\caption{Comparison of the sensitivities for the anomalous coupling $\Delta\kappa_{\gamma}$ in the pure leptonic, semi-leptonic and pure hadronic decay channels of the process $e^- p\,\rightarrow\,e^-\gamma^*\gamma^* p\,\rightarrow\,e^-W^+W^-p$ at the LHeC (left) and the FCC-he (right) with center-of-mass energies $\sqrt{s}=1.30, 1.98, 3.46, 5.29$ TeV and integrated luminosities ${\cal L}_{\text{int}}=10, 30, 50, 100, 500, 1000, 2000$ fb$^{-1}$.     
\label{fig4}}
\end{figure}

\begin{figure}[H]
\centering
\includegraphics[scale=0.8]{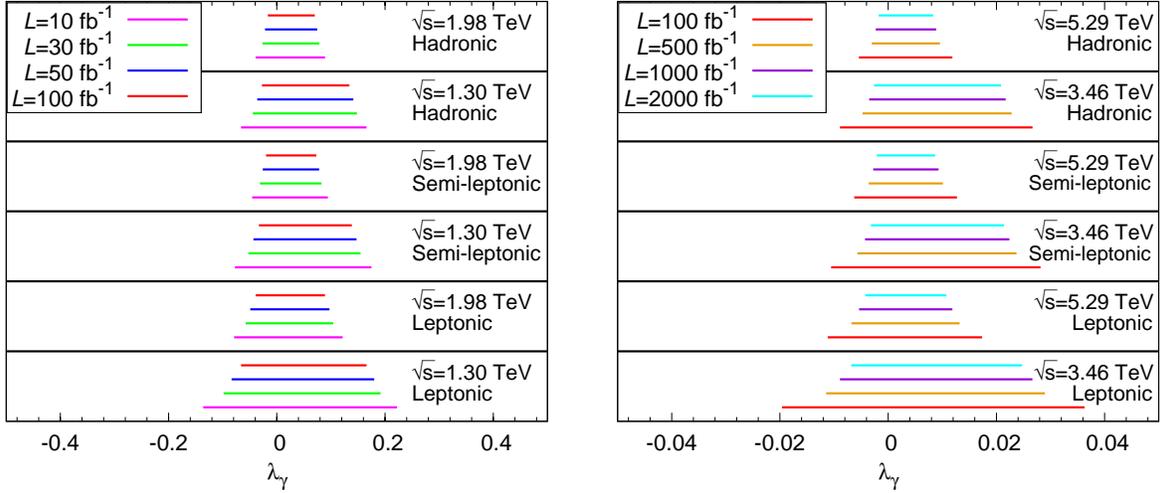}
\caption{Same as in Fig.~\ref{fig4}, but for the anomalous coupling $\lambda_{\gamma}$.     
\label{fig5}}
\end{figure}

In Figs.~\ref{fig4}-\ref{fig5}, the effects of the center-of-mass energy, integrated luminosity and decay channels on the bounds are examined. Thus, it is possible to compare the sensitivity of the anomalous couplings $\Delta\kappa_{\gamma}$ and $\lambda_{\gamma}$ between different collider options. It is observed that the sensitivity of the anomalous couplings $\Delta\kappa_{\gamma}$ and $\lambda_{\gamma}$ increases with the increase in the values of center-of-mass energy and integrated luminosity. Moreover, in proportion to the branching ratios of the decay channels of the process, the most sensitive bounds are at pure hadronic decay channels whereas the bounds at pure leptonic decay channels are less sensitive. According to Tables~\ref{tab3}-\ref{tab6}, the FCC-he collider with $\sqrt{s}=5.29$ TeV and ${\cal L}_{\text{int}}=2000$ fb$^{-1}$ for pure hadronic decay channel has the most sensitive bounds on the anomalous couplings $\Delta\kappa_{\gamma}$ and $\lambda_{\gamma}$. These bounds are given by;

\begin{eqnarray}
\label{eq.21} 
\Delta\kappa_{\gamma}=[-0.00208; 0.00206]\,,
\end{eqnarray}
\begin{eqnarray}
\label{eq.22} 
\lambda_{\gamma}=[-0.00174; 0.00834]\,.
\end{eqnarray}

\begin{table}[H]
\caption{95\% C.L. bounds on the anomalous $\Delta\kappa_{\gamma}$ coupling in the pure leptonic, semi-leptonic and pure hadronic decay channels of the process $e^- p\,\rightarrow\,e^-\gamma^*\gamma^* p\,\rightarrow\,e^-W^+W^-p$ at the LHeC colliders for the systematic errors and the integrated luminosities.}
\label{tab3}
\centering
\begin{ruledtabular}
\begin{tabular}{|c|ccccc|}
\multicolumn{3}{c}{} & \multicolumn{3}{c}{$\Delta\kappa_{\gamma}$} \\
\hline
Colliders & Channels & ${\cal L}_{\text{int}}$ (fb$^{-1}$) & $\delta_{sys}=0\%$ & $\delta_{sys}=3\%$ & $\delta_{sys}=5\%$ \\ 
\hline \hline
\multirow{12}{*}{LHeC-1} 
 & \multirow{4}{*}{Pure leptonic}  
 & $10$ & [-0.43440; 0.18913] & [-0.44032; 0.18998] & [-0.45117; 0.19148]\\ 
 & & $30$ & [-0.17737; 0.11952] & [-0.18129; 0.12124] & [-0.18823; 0.12423]\\ 
 & & $50$ & [-0.12868; 0.09553] & [-0.13302; 0.09788] & [-0.14060; 0.10187]\\ 
 & & $100$ & [-0.08592; 0.06986] & [-0.09127; 0.07335] & [-0.10035; 0.07907]\\ \cline{2-6} 
  & \multirow{4}{*}{Semi-leptonic}  
 & $10$ & [-0.11193; 0.08605] & [-0.11656; 0.08874] & [-0.12458; 0.09327]\\ 
 & & $30$ & [-0.06030; 0.05195] & [-0.06702; 0.05686] & [-0.07791; 0.06449]\\ 
 & & $50$ & [-0.04582; 0.04084] & [-0.05392; 0.04715] & [-0.06639; 0.05640]\\ 
 & & $100$ & [-0.03180; 0.02932] & [-0.04217; 0.03791] & [-0.05669; 0.04925]\\ \cline{2-6}  
  & \multirow{4}{*}{Pure hadronic}  
 & $10$ & [-0.08602; 0.06993] & [-0.09137; 0.07341] & [-0.10045; 0.07913]\\ 
 & & $30$ & [-0.04715; 0.04189] & [-0.05509; 0.04804] & [-0.06739; 0.05712]\\ 
 & & $50$ & [-0.03598; 0.03284] & [-0.04554; 0.04061] & [-0.05939; 0.05127]\\ 
 & & $100$ & [-0.02507; 0.02351] & [-0.03712; 0.03379] & [-0.05280; 0.04629]\\ \hline\hline  
\multirow{12}{*}{LHeC-2} 
 & \multirow{4}{*}{Pure leptonic}  
 & $10$ & [-0.24616; 0.13671] & [-0.25105; 0.13807] & [-0.25987; 0.14044]\\ 
 & & $30$ & [-0.11578; 0.08520] & [-0.12084; 0.08787] & [-0.12965; 0.09237]\\ 
 & & $50$ & [-0.08557; 0.06774] & [-0.09141; 0.07132] & [-0.10131; 0.07717]\\ 
 & & $100$ & [-0.05798; 0.04925] & [-0.06534; 0.05444] & [-0.07718; 0.06239]\\ \cline{2-6}  
  & \multirow{4}{*}{Semi-leptonic}  
 & $10$ & [-0.07488; 0.06088] & [-0.08117; 0.06496] & [-0.09168; 0.07149]\\ 
 & & $30$ & [-0.04103; 0.03647] & [-0.05024; 0.04355] & [-0.06405; 0.05354]\\ 
 & & $50$ & [-0.03131; 0.02858] & [-0.04227; 0.03744] & [-0.05755; 0.04894]\\ 
 & & $100$ & [-0.02182; 0.02046] & [-0.03540; 0.03195] & [-0.05231; 0.04510]\\ \cline{2-6}  
  & \multirow{4}{*}{Pure hadronic}  
 & $10$ & [-0.05805; 0.04930] & [-0.06540; 0.05448] & [-0.07724; 0.06243]\\ 
 & & $30$ & [-0.03221; 0.02933] & [-0.04297; 0.03799] & [-0.05811; 0.04934]\\ 
 & & $50$ & [-0.02466; 0.02294] & [-0.03733; 0.03351] & [-0.05374; 0.04616]\\ 
 & & $100$ & [-0.01724; 0.01638] & [-0.03258; 0.02964] & [-0.05029; 0.04359]\\ 
\end{tabular}
\end{ruledtabular}
\end{table}

\begin{table}[H]
\caption{Same as in Table~\ref{tab3}, but for the FCC-he colliders.}
\label{tab4}
\centering
\begin{ruledtabular}
\begin{tabular}{|c|ccccc|}
\multicolumn{3}{c}{} & \multicolumn{3}{c}{$\Delta\kappa_{\gamma}$} \\
\hline
Colliders & Channels & ${\cal L}_{\text{int}}$ (fb$^{-1}$) & $\delta_{sys}=0\%$ & $\delta_{sys}=3\%$ & $\delta_{sys}=5\%$ \\ 
\hline \hline
\multirow{12}{*}{FCC-he-1} 
 & \multirow{4}{*}{Pure leptonic}  
 & $100$ & [-0.03967; 0.03428] & [-0.04994; 0.04167] & [-0.06535; 0.05183]\\ 
 & & $500$ & [-0.01695; 0.01589] & [-0.03324; 0.02938] & [-0.05519; 0.04323]\\ 
 & & $1000$ & [-0.01186; 0.01133] & [-0.03069; 0.02737] & [-0.05340; 0.04199]\\ 
 & & $2000$ & [-0.00833; 0.00806] & [-0.02935; 0.02630] & [-0.05248; 0.04136]\\ \cline{2-6} 
  & \multirow{4}{*}{Semi-leptonic}  
 & $100$ & [-0.01502; 0.01418] & [-0.03219; 0.02856] & [-0.05296; 0.04272]\\ 
 & & $500$ & [-0.00660; 0.00644] & [-0.03084; 0.02589] & [-0.05135; 0.04112]\\ 
 & & $1000$ & [-0.00465; 0.00457] & [-0.03018; 0.02553] & [-0.05115; 0.04092]\\ 
 & & $2000$ & [-0.00328; 0.00324] & [-0.02908; 0.02535] & [-0.05105; 0.04082]\\ \cline{2-6}  
  & \multirow{4}{*}{Pure hadronic}  
 & $100$ & [-0.01188; 0.01134] & [-0.03069; 0.02737] & [-0.05040; 0.04200]\\ 
 & & $500$ & [-0.00524; 0.00513] & [-0.02852; 0.02563] & [-0.04893; 0.04097]\\ 
 & & $1000$ & [-0.00369; 0.00364] & [-0.02824; 0.02540] & [-0.04875; 0.04084]\\ 
 & & $2000$ & [-0.00260; 0.00258] & [-0.02809; 0.02529] & [-0.04865; 0.04078]\\ \hline\hline  
\multirow{12}{*}{FCC-he-2} 
 & \multirow{4}{*}{Pure leptonic}  
 & $100$ & [-0.03191; 0.02738] & [-0.04464; 0.03621] & [-0.06307; 0.04734]\\ 
 & & $500$ & [-0.01359; 0.01270] & [-0.03240; 0.02774] & [-0.05348; 0.04178]\\ 
 & & $1000$ & [-0.00951; 0.00906] & [-0.03064; 0.02644] & [-0.05222; 0.04101]\\ 
 & & $2000$ & [-0.00667; 0.00645] & [-0.02923; 0.02576] & [-0.05159; 0.04062]\\ \cline{2-6}  
  & \multirow{4}{*}{Semi-leptonic}  
 & $100$ & [-0.01204; 0.01134] & [-0.03168; 0.02721] & [-0.05145; 0.04146]\\ 
 & & $500$ & [-0.00529; 0.00515] & [-0.02940; 0.02551] & [-0.04915; 0.04048]\\ 
 & & $1000$ & [-0.00372; 0.00365] & [-0.02910; 0.02529] & [-0.04885; 0.04035]\\ 
 & & $2000$ & [-0.00262; 0.00259] & [-0.02895; 0.02518] & [-0.04871; 0.04029]\\ \cline{2-6}  
  & \multirow{4}{*}{Pure hadronic}  
 & $100$ & [-0.00952; 0.00907] & [-0.03035; 0.02645] & [-0.05023; 0.04101]\\ 
 & & $500$ & [-0.00420; 0.00411] & [-0.02818; 0.02535] & [-0.04720; 0.04039]\\ 
 & & $1000$ & [-0.00296; 0.00291] & [-0.02799; 0.02521] & [-0.04708; 0.04031]\\ 
 & & $2000$ & [-0.00208; 0.00206] & [-0.02790; 0.02514] & [-0.04701; 0.04027]\\ 
\end{tabular}
\end{ruledtabular}
\end{table}

\begin{table}[H]
\caption{95\% C.L. bounds on the anomalous $\lambda_{\gamma}$ coupling in the pure leptonic, semi-leptonic and pure hadronic decay channels of the process $e^- p\,\rightarrow\,e^-\gamma^*\gamma^* p\,\rightarrow\,e^-W^+W^-p$ at the LHeC colliders for the systematic errors and the integrated luminosities.}
\label{tab5}
\centering
\begin{ruledtabular}
\begin{tabular}{|c|ccccc|}
\multicolumn{3}{c}{} & \multicolumn{3}{c}{$\lambda_{\gamma}$} \\
\hline
Colliders & Channels & ${\cal L}_{\text{int}}$ (fb$^{-1}$) & $\delta_{sys}=0\%$ & $\delta_{sys}=3\%$ & $\delta_{sys}=5\%$ \\ 
\hline \hline
\multirow{12}{*}{LHeC-1} 
 & \multirow{4}{*}{Pure leptonic}  
 & $10$ & [-0.13597; 0.22215] & [-0.13639; 0.22250] & [-0.13713; 0.22310]\\ 
 & & $30$ & [-0.09845; 0.19180] & [-0.09947; 0.19262] & [-0.10122; 0.19403]\\ 
 & & $50$ & [-0.08362; 0.17986] & [-0.08513; 0.18108] & [-0.08766; 0.18312]\\ 
 & & $100$ & [-0.06611; 0.16574] & [-0.06861; 0.16776] & [-0.07263; 0.17100]\\ \cline{2-6} 
  & \multirow{4}{*}{Semi-leptonic}  
 & $10$ & [-0.07737; 0.17483] & [-0.07917; 0.17628] & [-0.08215; 0.17868]\\ 
 & & $30$ & [-0.05253; 0.15472] & [-0.05639; 0.15786] & [-0.06218; 0.16256]\\ 
 & & $50$ & [-0.04333; 0.14719] & [-0.04864; 0.15154] & [-0.05603; 0.15757]\\ 
 & & $100$ & [-0.03297; 0.13864] & [-0.04079; 0.14511] & [-0.05036; 0.15295]\\ \cline{2-6}  
  & \multirow{4}{*}{Pure hadronic}  
 & $10$ & [-0.06616; 0.16578] & [-0.06866; 0.16780] & [-0.07267; 0.17104]\\ 
 & & $30$ & [-0.04424; 0.14794] & [-0.04937; 0.15214] & [-0.05660; 0.15803]\\ 
 & & $50$ & [-0.03624; 0.14135] & [-0.04314; 0.14704] & [-0.05199; 0.15428]\\ 
 & & $100$ & [-0.02734; 0.13396] & [-0.03711; 0.14207] & [-0.04794; 0.15097]\\ \hline\hline  
\multirow{12}{*}{LHeC-2} 
 & \multirow{4}{*}{Pure leptonic}  
 & $10$ & [-0.07886; 0.12166] & [-0.07937; 0.12208] & [-0.08024; 0.12281]\\ 
 & & $30$ & [-0.05760; 0.10416] & [-0.05882; 0.10516] & [-0.06085; 0.10682]\\ 
 & & $50$ & [-0.04915; 0.09723] & [-0.05096; 0.09871] & [-0.05381; 0.10105]\\ 
 & & $100$ & [-0.03914; 0.08899] & [-0.04208; 0.09142] & [-0.04638; 0.09496]\\ \cline{2-6}  
  & \multirow{4}{*}{Semi-leptonic}  
 & $10$ & [-0.04559; 0.09430] & [-0.04773; 0.09606] & [-0.05104; 0.09877]\\ 
 & & $30$ & [-0.03132; 0.08252] & [-0.03576; 0.08620] & [-0.04158; 0.09101]\\ 
 & & $50$ & [-0.02598; 0.07807] & [-0.03194; 0.08304] & [-0.03895; 0.08884]\\ 
 & & $100$ & [-0.01992; 0.07298] & [-0.02831; 0.08002] & [-0.03669; 0.08697]\\ \cline{2-6}  
  & \multirow{4}{*}{Pure hadronic}  
 & $10$ & [-0.03917; 0.08902] & [-0.04211; 0.09144] & [-0.04640; 0.09497]\\ 
 & & $30$ & [-0.02651; 0.07851] & [-0.03229; 0.08333] & [-0.03918; 0.08903]\\ 
 & & $50$ & [-0.02184; 0.07460] & [-0.02936; 0.08090] & [-0.03732; 0.08749]\\ 
 & & $100$ & [-0.01660; 0.07016] & [-0.02671; 0.07869] & [-0.03577; 0.08622]\\ 
\end{tabular}
\end{ruledtabular}
\end{table}

\begin{table}[H]
\caption{Same as in Table~\ref{tab5}, but for the FCC-he colliders.}
\label{tab6}
\centering
\begin{ruledtabular}
\begin{tabular}{|c|ccccc|}
\multicolumn{3}{c}{} & \multicolumn{3}{c}{$\lambda_{\gamma}$} \\
\hline
Colliders & Channels & ${\cal L}_{\text{int}}$ (fb$^{-1}$) & $\delta_{sys}=0\%$ & $\delta_{sys}=3\%$ & $\delta_{sys}=5\%$ \\ 
\hline \hline
\multirow{12}{*}{FCC-he-1} 
 & \multirow{4}{*}{Pure leptonic}  
 & $100$ & [-0.01964; 0.03629] & [-0.02236; 0.03869] & [-0.02575; 0.04169]\\ 
 & & $500$ & [-0.01146; 0.02898] & [-0.01769; 0.03456] & [-0.02290; 0.03917]\\ 
 & & $1000$ & [-0.00890; 0.02666] & [-0.01686; 0.03382] & [-0.02247; 0.03879]\\ 
 & & $2000$ & [-0.00682; 0.02476] & [-0.01640; 0.03341] & [-0.02224; 0.03859]\\ \cline{2-6} 
  & \multirow{4}{*}{Semi-leptonic}  
 & $100$ & [-0.01053; 0.02815] & [-0.01736; 0.03426] & [-0.02272; 0.03901]\\ 
 & & $500$ & [-0.00569; 0.02372] & [-0.01622; 0.03326] & [-0.02216; 0.03852]\\ 
 & & $1000$ & [-0.00427; 0.02241] & [-0.01607; 0.03312] & [-0.02209; 0.03846]\\ 
 & & $2000$ & [-0.00318; 0.02139] & [-0.01599; 0.03305] & [-0.02205; 0.03842]\\ \cline{2-6}  
  & \multirow{4}{*}{Pure hadronic}  
 & $100$ & [-0.00890; 0.02667] & [-0.01686; 0.03382] & [-0.02247; 0.03879]\\ 
 & & $500$ & [-0.00472; 0.02282] & [-0.01611; 0.03315] & [-0.02211; 0.03847]\\ 
 & & $1000$ & [-0.00352; 0.02171] & [-0.01601; 0.03307] & [-0.02206; 0.03843]\\ 
 & & $2000$ & [-0.00260; 0.02085] & [-0.01596; 0.03302] & [-0.02204; 0.03841]\\ \hline\hline  
\multirow{12}{*}{FCC-he-2} 
 & \multirow{4}{*}{Pure leptonic}  
 & $100$ & [-0.01118; 0.01736] & [-0.01330; 0.01935] & [-0.01566; 0.02157]\\ 
 & & $500$ & [-0.00679; 0.01320] & [-0.01127; 0.01745] & [-0.01451; 0.02050]\\ 
 & & $1000$ & [-0.00538; 0.01186] & [-0.01093; 0.01713] & [-0.01435; 0.02034]\\ 
 & & $2000$ & [-0.00423; 0.01075] & [-0.01076; 0.01696] & [-0.01427; 0.02026]\\ \cline{2-6}  
  & \multirow{4}{*}{Semi-leptonic}  
 & $100$ & [-0.00628; 0.01272] & [-0.01113; 0.01732] & [-0.01445; 0.02043]\\ 
 & & $500$ & [-0.00358; 0.01012] & [-0.01069; 0.01690] & [-0.01424; 0.02024]\\ 
 & & $1000$ & [-0.00276; 0.00932] & [-0.01063; 0.01685] & [-0.01421; 0.02021]\\ 
 & & $2000$ & [-0.00210; 0.00868] & [-0.01060; 0.01682] & [-0.01420; 0.02020]\\ \cline{2-6}  
  & \multirow{4}{*}{Pure hadronic}  
 & $100$ & [-0.00539; 0.01186] & [-0.01093; 0.01713] & [-0.01435; 0.02034]\\ 
 & & $500$ & [-0.00302; 0.00958] & [-0.01065; 0.01686] & [-0.01422; 0.02022]\\ 
 & & $1000$ & [-0.00230; 0.00889] & [-0.01061; 0.01683] & [-0.01420; 0.02020]\\ 
 & & $2000$ & [-0.00174; 0.00834] & [-0.01059; 0.01681] & [-0.01419; 0.02019]\\ 
\end{tabular}
\end{ruledtabular}
\end{table}

\section{Conclusions}

We have performed studies on the sensitivity analysis of the anomalous $W^+W^-\gamma$ couplings in the process $e^- p\,\rightarrow\,e^-\gamma^*\gamma^* p\,\rightarrow\,e^-W^+W^-p$ with pure hadronic, semi-leptonic and pure leptonic decay channels at the LHeC and the FCC-he. It can be seen from Tables~\ref{tab3}-\ref{tab6} that the pure hadronic channel with the largest branching ratio of W-pair has the highest sensitivity compared to other channels with the same parameters and the pure leptonic channel with the smallest branching ratio of W-pair has the lowest sensitivity compared to other channels with the same parameters. In this paper, the most sensitive bounds of the anomalous couplings $\Delta\kappa_{\gamma}$ and $\lambda_{\gamma}$ are given in Eqs.~(\ref{eq.21})-(\ref{eq.22}). Comparing these bounds with the $\Delta\kappa_{\gamma}=[-0.0275; 0.0286]$ and $\lambda_{\gamma}=[-0.0065; 0.0066]$ limits at the CMS Collaboration which has the current best experimental limits for the anomalous couplings $\Delta\kappa_{\gamma}$ and $\lambda_{\gamma}$ in Table~\ref{tab1}, reveals the potential of future $ep$ colliders. The sensitivity of the $\Delta\kappa_{\gamma}$ coupling in Eq.~(\ref{eq.21}) is about 14 times better than the sensitivity of the experimental limit at the CMS Collaboration in Table~\ref{tab1}. Likewise, the sensitivity of the $\lambda_{\gamma}$ coupling in Eq.~(\ref{eq.22}) is about 2 times better than the sensitivity of the experimental limit at the CMS Collaboration in Table~\ref{tab1}.

As a result, the process $e^- p\,\rightarrow\,e^-\gamma^*\gamma^* p\,\rightarrow\,e^-W^+W^-p$ is of great importance for measurement of the anomalous $W^+W^-\gamma$ couplings and it is obvious that the future $ep$ colliders which have a clean environment will also contribute to the LHC's physics program about the exploration of new physics beyond the SM.

\end{document}